\documentclass[referee]{mn2e}
\usepackage[dvips]{graphicx}
\usepackage{epsfig}

\usepackage{txfonts}

\title[Bulk viscosity of Mixed nucleon-hyperon-quark Matter in Neutron stars]{Bulk viscosity of Mixed nucleon-hyperon-quark Matter in Neutron stars}

\author[Na-Na Pan , Xiao-Ping Zheng  and Jia-Rong Li ]{Na-Na Pan$^{1}$, Xiao-Ping Zheng$^{1}$\thanks{E-mail:zhxp@phy.ccnu.edu.cn} and Jia-Rong Li$^{2}$
\\ 1.Institute of Astrophysics, Huazhong Normal
University, Wuhan 430079, China
\\ 2.Institute of Particle Physics, Huazhong Normal
University, Wuhan 430079, China}

\date{Accepted 0000, Received 0000}
\pagerange{\pageref{firstpage}--\pageref{lastpage}} \pubyear{0000}

\begin{document}
\label{firstpage} \maketitle

\begin{abstract}
We calculate the coefficient of bulk viscosity by considering the
non-leptonic weak interactions in the cores of hybrid stars with
both hyperons and quarks. We first  determine the dependence of
the production rate of neutrons on the reaction rate of quarks in
the non-leptonic processes, that is
$\Gamma_{n}=K_{s}\Gamma_{s}+\Gamma_{\Lambda}+2\Gamma_{\Sigma^{-}}$.
The conversion rate, $K_{s}$ in our scenario is a complicated
function of baryon number density. We also consider medium effect
of quark matter on bulk viscosity. Using these results, we
estimate the limiting rotation of the hybrid stars, which may
suppress the r-mode instability more effectively. Hybrid stars
should be the candidates for  the extremely rapid rotators .
\end{abstract}


\section{Introduction}
It has been recognized that  weak interaction processes in neutron
star matter contribute to the bulk viscosity(Sawyer \&
Soui\cite{1}; Sawyer \cite{2}; Haensel \& Schaeffer \cite{3}).
When only nucleons and leptons exist in the interior of neutron
stars, the bulk viscosity results from Urca processes.   Some
 exotic particles, such as hyperons and quarks, can be produced as
the density inside neutron stars increases (Weber \cite{4};
Glendenning et al. \cite{5}). The non-leptonic processes involving
both hyperons and quarks  lead to larger coefficient of bulk
viscosity of dense nuclear matter(Jones \cite{6}; Wang \& Lu
\cite{8}). Therefore, the calculations of the bulk viscosity
induced by non-leptonic reactions have received considerable
attention in the past few years(Sawyer \cite{9}; Madsen \cite{10};
Goyal et al. \cite{11}; Haensel \& Levenfish \cite{12}; Haensel et
al. \cite{34};Lindblom \& Owen \cite{13}; Zheng et al.\cite{14};
Zheng et al.\cite{15}; Zheng et al.\cite{16})since the discovery
by Andersson, Friedman and Morsink of r-mode instability in
neutron stars(Andersson \cite{17}; Friedman \& Morsink \cite{18}).
Actually, neutron stars can be constituted by a larger variety of
particles other than just neutrons and protons or pure quark
matter. The mixed phases, with nucleon-hyperon, nucleon-quark or
nucleon-hyperon-quark matter, may be contained in the interior of
neutron stars(Glendenning \cite{19}; Glendenning \cite{20}). The
bulk viscosities of mixed nucleon-hyperon matter and nucleon-quark
matter have been discussed by some authors so far(Lindblom \& Owen
\cite{13}; Drago et al. \cite{21}; Nayyar \& Owen \cite{22}).
However, the calculations are
 reduced to single reaction case since only one type of
reaction is involved, either hadron or quark non-leptonic process.
The purpose of this paper is to investigate the general case
thoroughly. We focus on mixed nucleon-hyperon-quark matter to
evaluate its bulk viscosity by
 considering the reactions involving both hyperons and quarks.
According to Lindblom \& Owen (LO) formalism (Lindblom \& Owen
\cite{13};  Nayyar \& Owen \cite{22}), the bulk viscosity depends
on the relaxation time  governed by the hyperon and quark
reactions. Following the LO approach, we first convert reaction
rates for hyperons and quarks into the production rate of neutrons
per unit volume in order to give the relaxation time of the
system, and then simulate the bulk viscosity of the mixed phase
using the rates. In our study, the conversion rate from quarks to
neutrons varies with a variable baryon number density, and we will
see below that the numerical solution is necessary. Our study is
under the neutron stars models based on $Ghosh-Phatak-Sahu$
$(GPS)$ equations of state(Ghosh et al.\cite{23}) (EOS) for hadron
matter and $SGT$ model (Schertler, Greiner \& Thoma \cite{24}) for
quark matter in medium. At the high density, $GPS$ EOSs are
favorable for the equilibrium state to include a sequence of
hyperons. Our model also concerns a $Baym-Pethick-Sutherland $
$(BPS)$(Baym et al.\cite{25}) crust. Our analysis shows that there
exist two instability windows for a neutron star with strangeness.
The low-temperature window indicates the possibility of existence
of submillisecond pulsars.

The organization of the rest of this paper is as follows. In
section 2, we provide details of structures in stars which we are
studying by taking into account a first-order deconfinement phase
transition into quarks from hadrons in the cores of the stars.
Then in section 3,we employ LO approach to derive the bulk
viscosity of mixed nucleon-hyperon-quark matter, including
numerical aspects of the evaluations of the viscosity coefficient.
Finally in section 4, we work out r-mode instability window based
on the competition between increasing of gravitational radiation
and damping of viscosities.

\section{Equation of state}

In this paper, we consider the  compact stars inside which the
deconfinement transition  occurs at high density. Concerning the
hadronic phase (HP), we use the $BPS$ EOS  at subnuclear densities
for the crust of the star, and the  $GPS$ EOS for supranuclear
density in the framework of the relativistic self-consistent mean
field theory (Glendenning \cite{19}). In this theory, $\Sigma^{-}$
and $\Lambda$ hyperons are included for nuclear densities, where
nucleons interact through the nuclear force mediated by the
exchange of isoscalar and isovector mesons $(\sigma,\omega,\rho)$.
The model parameters used in this paper are arranged in table 1.
These two EOSs are matched at
$\epsilon\approx10^{14}g/cm^{3}\approx0.4\epsilon_{0}$, here
$\epsilon_{0}=140Mev/fm^{3}\approx2.5\times10^{14}g/cm^{3}$ is the
saturation density of nucleon, above which we allow the HP to
undergo a first order phase transition to a deconfined quark
matter phase (QP) that is described with the effective mass bag
model considering medium effect represent by the coupling constant
$g$ for strong action (Schertler, Greiner \& Thoma \cite{24}).
This phase transition makes it possible that the occurrence of a
mixed hadron-quark phase (MP) in a finite density range inside
compact stars. The relative constraints of various species in the
MP are determined at each density by imposing $\beta$-equilibrium,
Gibbs condition and global charge neutrality. At the high density
of interest to us, these equilibrium constraints are
\begin{equation}
\mu_{n}=\mu_{u}+2\mu_{d},
\end{equation}
\begin{equation}
\mu_{p}=2\mu_{u}+\mu_{d},
\end{equation}
\begin{equation}
\mu_{\mu}=\mu_{e},
\end{equation}
\begin{equation}
\mu_{\Sigma^{-}}=\mu_{n}+\mu_{e},
\end{equation}
\begin{equation}
\mu_{\Lambda}=\mu_{n},
\end{equation}
\begin{equation}
\mu_{s}=\mu_{d},
\end{equation}
\begin{equation}
\mu_{u}=\mu_{d}-\mu_{e},
\end{equation}
\begin{equation}
P_{HP}(\mu_{n},\mu_{e})=P_{QP}(\mu_{n},\mu_{e}),
\end{equation}
\begin{equation}
(1-\chi)(n_{p}-n_{\Sigma^{-}}-n_{HP,
e}-n_{\mu})=\chi(-\frac{2}{3}n_{u}+\frac{1}{3}n_{d}+\frac{1}{3}n_{s}+n_{QP,
e}).
\end{equation}

\begin{figure}
\includegraphics[width=0.9\textwidth]{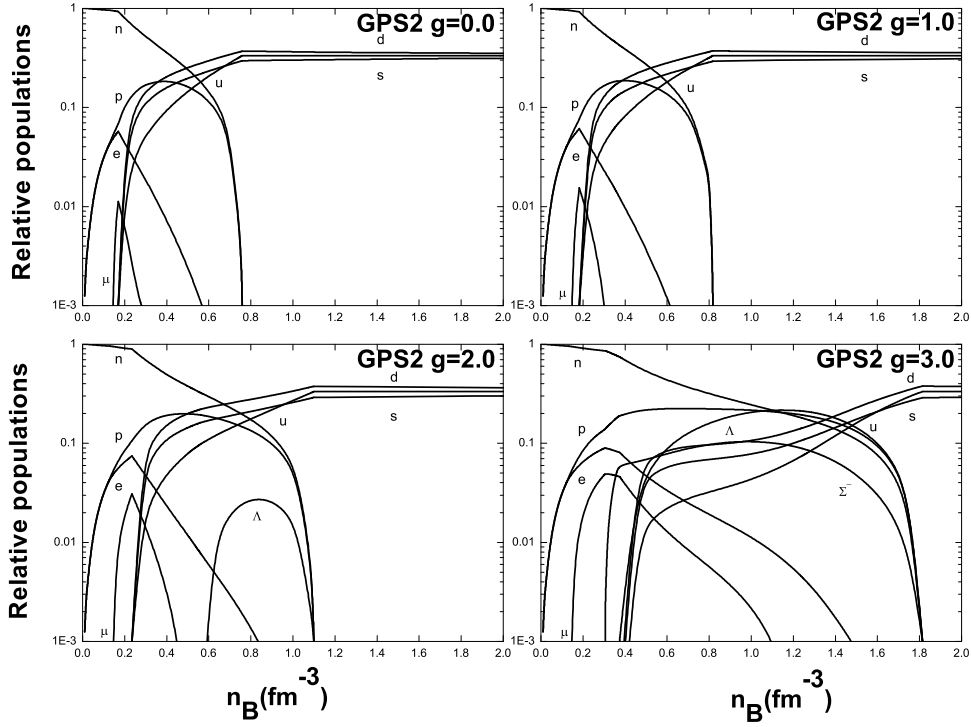}
\caption{Influences of medium effect on the relative particles
populations as functions of baryon number densities in the hybrid
stars. Note that under the condition $g=1.0$, $\Lambda$ does also
exist in the MP, but its population is small compared with
others', so we don't show it here. Here they are the relative
populations of baryon number densities for u,d,s quark only. For
all QP EOSs, we choose the bag constant $B^{1/4}=170.0 Mev$  and
the mass of the s quark $m_{s}=150.0 Mev$ in calculations.}
\label{GPS fixed population}
\end{figure}

\begin{figure}
\centering
\includegraphics[width=0.6\textwidth]{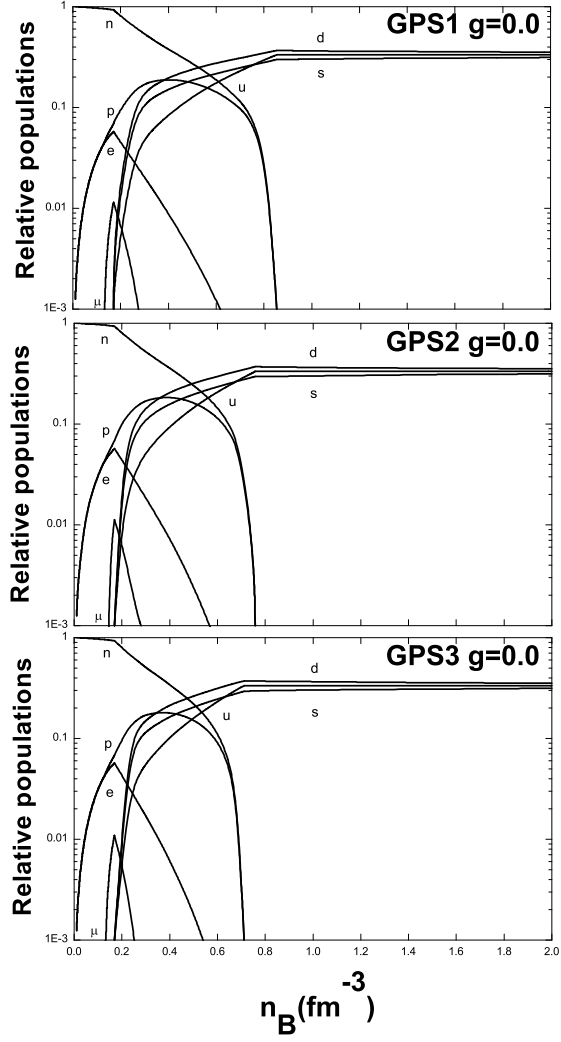}
\caption{Influences of hadron EOSs on relative particles
populations as functions of baryon number densities in the hybrid
stars. Under the condition $GPS1$, $\Lambda$ also exists in the
MP, but we don't show it here as its population is small compared
with others'. Parameters of QP EOSs are the same with the Figure 1
in calculations.} \label{g fixed population}
\end{figure}

\begin{figure}
\centering
\includegraphics[width=0.8\textwidth]{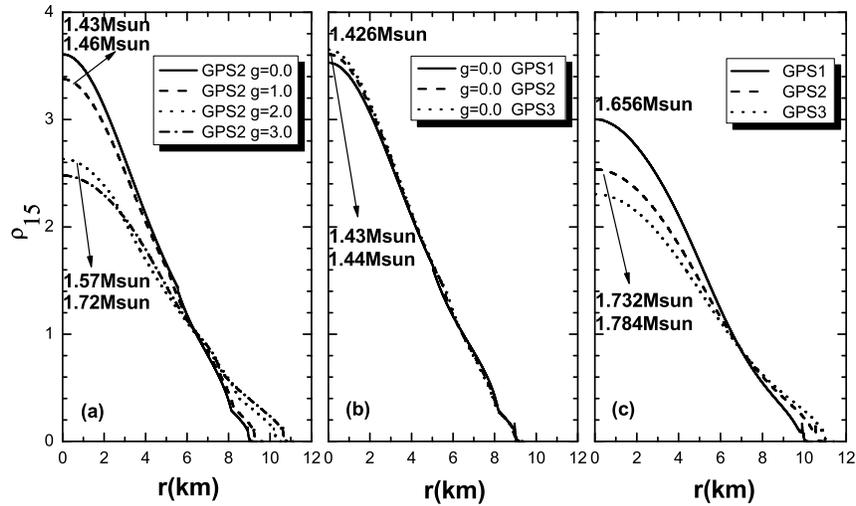}
\caption{Structures of maximum masses hybrid stars (a) (b) using
different EOSs discussed in the Figure 1 and 2, and the
corresponding hyperon stars (c) for same hadron EOSs with case (b)
 : the relations
between total energy densities(in units of $10^{15}g/cm^{3}$) and
the distances from the center of the stars .} \label{structure}
\end{figure}

Here $\mu_{i},n_{i}$ are the chemical potential and the number
density of the $i^{th}$ species respectively, $P_{i}$ is the
pressure of the $i^{th}$ phase, and $\chi$ is the fraction of the
quark matter in MP. Solving these constraints, we can give the
compositions of the equilibrium matter in hybrid stars. Figure
\ref{GPS fixed population} and \ref{g fixed population}  show some
solutions for different parameters. Evidently, they contain a
significant number of hyperons and strange quark matter in the
highest density portion of their cores, and the additional
hyperons appear as the coupling among quarks increases. Using the
Tolman-Oppenheimer-Volkoff (TOV) equation, we obtain the
structures of some hybrid stars with maximum masses for a few EOSs
shown in Figure \ref{structure}, where we also plot the structures
of hyperon stars for the same $GPS$ EOSs (Figure
\ref{structure}(c)) as comparison. The EOSs of the stars are
softened when deconfinement phase transition happens, but the
structures of the stars are not sensitive to the changes of $GPS$
EOSs unlike hyperon stars, while medium effect in quark matter
does significantly change the stellar masses and radii. As is
known, the Hulse-Taylor pulsar has the measured mass of 1.44
$M_{sun}$ (Taylor \& Weisberg \cite{35}). Strictly speaking, the
EOSs models producing the maximum mass smaller than the measured
value should be rejected. We, however, take the lower values
around 1.44 $M_{sun}$ to compare with $GPS$ models (hyperon
stars), since the small difference of the masses can't
 change our results significantly below.

\section{Bulk viscosity}
Bulk viscosity is the dissipative process in which the macroscopic
compression or expansion of a fluid element is converted to
heat(Landau \& Lifshitz \cite{26}). The dissipation due to bulk
viscosity is carried out via  microscopic reactions which come
from weak interactions in compact star matter. In general, there
are two types of reactions in dense nuclear matter. One produces
bulk viscosity because its timescales are comparable with the
period of the perturbation, the other puts constraints on the
variations of  different particles' densities. As in most cases,
the volume per unit mass variation and the chemical potential
imbalance due to the oscillation in the star are so small that it
is the linear part that contributes most to the viscosity, so  we
will use the relaxation time approximation method(Lindblom \& Owen
\cite{13}; Drago et al.\cite{21}; Landau \& Lifshitz \cite{26}) to
compute the bulk viscosity of dense nuclear matter.

Here we are   interested in the case that hyperons and quarks are
all present in the mixed phase. It is thought that the
non-leptonic processes are dominating over the bulk viscosity of
nuclear matter  with strangeness (Jones \cite{6}; Lindblom \& Owen
\cite{13}; Drago et al.\cite{21}). In the initial works(Drago et
al.\cite{21}), one estimated the results arising from a single
reaction, assuming either an absence of hyperon or quark pairing
in mixed hadron-quark phase. That treatment was for simplification
in the calculations. In fact, the processes that produce bulk
viscosity of the mixed system are
\begin{equation}\label{en21}
u+d\longleftrightarrow u+s
\end{equation}
for strange quark matter, and
\begin{equation}\label{en220}
n+n\longleftrightarrow p+\Sigma^{-},
\end{equation}
\begin{equation}\label{en22}
n+p\longleftrightarrow p+\Lambda
\end{equation}
for hyperon matter.

Accordingly, the processes which put constraints on the variations
of the densities of different particles come from the melting
reactions of nucleons
\begin{equation}\label{en23}
p\longleftrightarrow 2u+d,
\end{equation}
\begin{equation}\label{en24}
n\longleftrightarrow u+2d,
\end{equation}
\begin{equation}\label{en240}
n+\Lambda\longleftrightarrow p+\Sigma^{-},
\end{equation}
and hyperons
\begin{equation}\label{en25}
\Lambda\longleftrightarrow u+d+s
\end{equation}
into free quarks.

In the mixed phase, since the degrees of freedom, for example, the
number densities of various baryons, are related to each other
even out of thermodynamic equilibrium by constraints such as
conservation of baryon number,  we could express all of the
perturbed quantities in terms of a single one.  We here choose the
number density of neutrons $n_{n}$ as our primary variable. Since
the three reactions (\ref{en21})(\ref{en220})(\ref{en22}) all
contribute to the change of neutron numbers, we express the
production rate of neutrons per unit volume as
\begin{equation}\label{en160000}
\Gamma_{n}=K_{s}\Gamma_{s}+K_{\Lambda}\Gamma_{\Lambda}
+K_{\Sigma^{-}}\Gamma_{\Sigma^{-}},
\end{equation}
where $\Gamma_{s}$, $\Gamma_{\Lambda}$, $\Gamma_{\Sigma^{-}}$
indicate the reaction rates of processes (\ref{en21}),
(\ref{en220}) and (\ref{en22}) respectively, and $K_{s}$,
$K_{\Lambda}$, $K_{\Sigma^{-}}$ are the rates converting reaction
rates of each exotic particles, s quark, $\Lambda$ and
$\Sigma^{-}$ hyperon, into the production rate  of neutrons. For
fast processes, we know that their contributions to the relaxation
time of the system could be omitted compared with the slow
processes, so we may easily know that $K_{\Lambda}=1$ and
$K_{\Sigma^{-}}=2$ by the reactions (\ref{en220}) and
(\ref{en22}). The determination of $K_{s}$ is  a slightly
complicated matter. Its treatment will be given later. Obviously,
if the system is perturbed, the fraction of the quark matter in
the mixed phase $\chi$ would be changed through reactions
(\ref{en23}), (\ref{en24}) and (\ref{en25}), and  the pressure of
the system would be  equilibrated again after the perturbation in
order to satisfy the mechanical equilibrium. Therefore we have
following constraints:

\begin{eqnarray}\label{en26}
0=(1-\chi)(\delta n_{n}+\delta n_{p}+\delta n_{\Sigma^{-}}+\delta
n_{\Lambda})+\chi(\delta n_{u}+\delta n_{d}+\delta
n_{s})/3\nonumber\\
+\delta \chi((n_{u}+n_{d}+n_{s})/3
-n_{n}-n_{p}-n_{\Sigma^{-}}-n_{\Lambda}),
\end{eqnarray}
\begin{eqnarray}\label{en27}
0=(1-\chi)(\delta n_{p}-\delta n_{\Sigma^{-}})+\chi(2\delta
n_{u}-\delta n_{d}-\delta n_{s})/3\nonumber\\
+\delta \chi((2n_{u}-n_{d}-n_{s})/3-n_{p}+n_{\Sigma^{-}}),
\end{eqnarray}
\begin{equation}\label{en28}
0=\sum_{{H}}p_{H}\delta n_{H}-\sum_{{Q}}p_{Q}\delta n_{Q},
\end{equation}
\begin{equation}\label{en29}
0=\alpha_{pn}\delta n_{n}+\alpha_{pp}\delta
n_{p}+\alpha_{p\Sigma^{-}}\delta
n_{\Sigma^{-}}+\alpha_{p\Lambda}\delta
n_{\Lambda}-2\alpha_{uu}\delta n_{u}-\alpha_{dd}\delta n_{d},
\end{equation}
\begin{equation}\label{en30}
0=\alpha_{nn}\delta n_{n}+\alpha_{np}\delta
n_{p}+\alpha_{n\Sigma^{-}}\delta
n_{\Sigma^{-}}+\alpha_{n\Lambda}\delta
n_{\Lambda}-\alpha_{uu}\delta n_{u}-2\alpha_{dd}\delta n_{d},
\end{equation}
\begin{equation}\label{en300}
0=\beta_{n}\delta n_{n}+\beta_{p}\delta
n_{p}+\beta_{\Sigma^{-}}\delta
n_{\Sigma^{-}}+\beta_{\Lambda}\delta n_{\Lambda},
\end{equation}
\begin{equation}\label{en31}
0=\alpha_{\Lambda n}\delta n_{n}+\alpha_{\Lambda p}\delta
n_{p}+\alpha_{\Lambda \Sigma^{-}}\delta
n_{\Sigma^{-}}+\alpha_{\Lambda\Lambda}\delta
n_{\Lambda}-\alpha_{uu}\delta n_{u}-\alpha_{dd}\delta
n_{d}-\alpha_{ss}\delta n_{s}.
\end{equation}
The first constraint (\ref{en26}) is the baryon number
conservation of the system. The second one (\ref{en27}) imposes
the conservation of electric charge assuming all leptonic reaction
rates are much smaller and neglected. The third constraint
(\ref{en28}) is related to the mechanical equilibrium. The last
four (\ref{en29})-(\ref{en31}) describe the equilibrium with
respect to reactions (\ref{en23})-(\ref{en25}). Noteworthily, when
the $\Sigma^{-}$ density vanishes in the system,  $\delta
n_{\Sigma^{-}}=0$, and the melting process (\ref{en240}) doesn't
exist, nor does the process (\ref{en220}).

Here we use
\begin{equation}
\alpha_{ij}=(\partial \mu_{i}/\partial n_{j})_{n_{k}, k\neq j},
\end{equation}
\begin{equation}
\beta_{i}=\alpha_{ni}+\alpha_{\Lambda
i}-\alpha_{pi}-\alpha_{\Sigma^{-}i},
\end{equation}
\begin{equation}
p_{i}=\partial P/\partial n_{i}
\end{equation}
for shortening. Then we could obtain the variations of various
particles' number densities $\delta n_{i}$ and the quark fraction
$\delta \chi$ in the form of $\delta n_{n}$.

\begin{figure}
\centering
\includegraphics[width=0.6\textwidth]{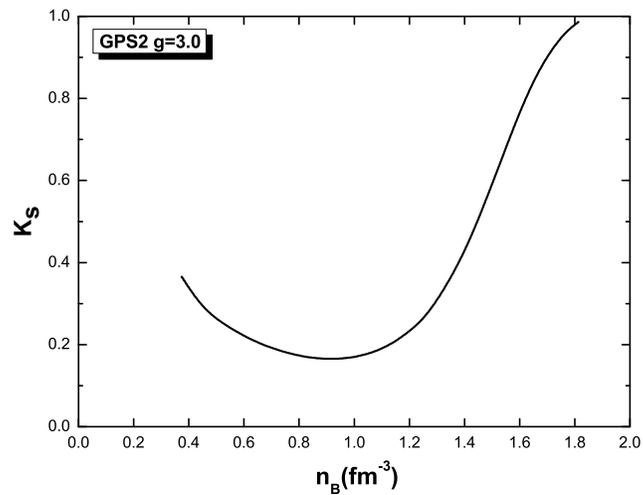}
\caption{Converting rate $K_{s}$ as a function of  baryon number
density in the mixed phase under the conditions $GPS2$ and
$g=3.0$.  } \label{Ks coefficient}
\end{figure}

In order to determine $K_{s}$, we need to describe the variation
of neutrons $\delta n_{n}$ following (\ref{en160000}) as

\begin{equation}\label{en16000}
-\delta n_{n}=K_{s}\delta n_{s}+\delta n_{\Lambda} +2\delta
n_{\Sigma^{-}}.
\end{equation}
Thus we have
\begin{equation}\label{en1600000}
K_{s}=\frac{-\delta n_{n}-\delta n_{\Lambda}-2\delta
n_{\Sigma^{-}}}{\delta n_{s}},
\end{equation}
which is a function of  baryon number density for given equation
of state under thermodynamic equilibrium in the mixed phase
system. Combining the constraints conditions to equation
(\ref{en1600000}) with the linear dependence of $\delta
n_{\Lambda}$, $\delta n_{\Sigma^{-}}$ and $\delta n_{s}$  on
$\delta n_{n}$, the $K_{s}$ can be determined. An example of the
numerical result is showed in the Figure \ref{Ks coefficient}. In
the extreme, if the system doesn't undergo the deconfinement phase
transition ($\delta n_{s}=0$) or is in the absence of hyperons
($\delta n_{\Lambda}=\delta n_{\Sigma^{-}}=0$), the calculation is
reduced to the situation of a single reaction(Lindblom \& Owen
\cite{13}; Drago et al.\cite{21}).

The relaxation time $\tau$ of the system due to these reactions
must be:
\begin{equation}\label{en1600}
\frac{1}{\tau}=(\frac{K_{s}\Gamma_{s}}{\delta\mu}+\frac{\Gamma_{\Lambda}}{\delta\mu}
+\frac{2\Gamma_{\Sigma^{-}}}{\delta\mu})\frac{\delta\mu}{\delta
n_{n}}.
\end{equation}
 Here $\Gamma_{s}$, $\Gamma_{\Lambda}$ and $\Gamma_{\Sigma^{-}}$  have already been  calculated by $Dai$ under  the high temperature limit
$(2\pi k T\gg \delta\mu)$ (Dai \& Lu \cite{27}) and $Lindblom$
$et$ $al.$ (Lindblom \& Owen \cite{13}; Nayyar \& Owen \cite{22})
separately. Considering the reactions
(\ref{en23})(\ref{en24})(\ref{en240})(\ref{en25}), the overall
chemical potential imbalance $\delta\mu$ is obtained through the
equilibrium of the system :
\begin{equation}
\delta \mu\equiv \delta \mu_{d}-\delta \mu_{s}= \delta
\mu_{n}-\delta \mu_{\Lambda}=2 \delta \mu_{n}-\delta
\mu_{p}-\delta \mu_{\Sigma^{-}},
\end{equation}
 which  could also be described in the form of  $\delta n_{n}$.

\begin{figure}
\centering
\includegraphics[width=0.6\textwidth]{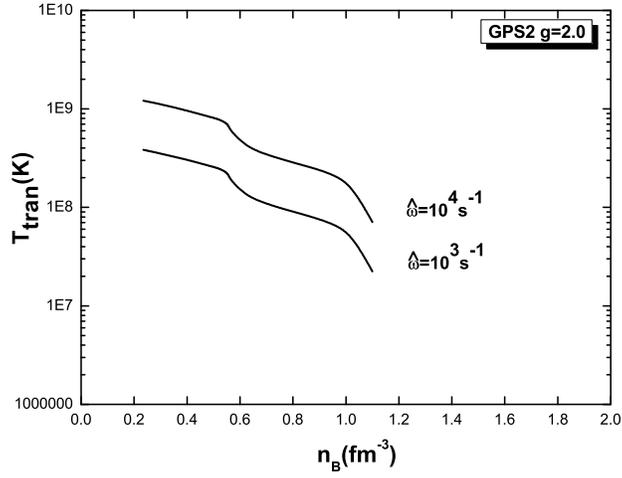}
\caption{Critical transition temperatures as functions of baryon
number densities and perturbation frequencies under the conditions
$GPS2$ and $g=2.0$.} \label{trantemperature}
\end{figure}

Now we use the LO formula
\begin{equation}\label{en16000}
Re\zeta=\frac{p(\gamma_{\infty}-\gamma_{0})\tau}{1+(\hat{\omega}\tau)^{2}}
\end{equation}
to evaluate the bulk viscosity. Thereinto,
 \begin{equation}
\gamma_{\infty}-\gamma_{0}\equiv
-\frac{n_{B}^{2}}{p}\frac{\partial p}{\partial
n_{n}}\frac{d\tilde{x}_{n}}{dn_{B}}.
\end{equation}
Here $\tilde{x}_{n}=n_{n}/n_{B}$ is the relative population of
neutrons in the equilibrium state of the mixed phase calculated in
the part 2. $\hat{\omega}$ is the perturbation frequency in
compact stars, which is typically $10^{3}\sim10^{4}s^{-1}$. We
have chosen $\hat{\omega}=10^{4}s^{-1}$ for calculation . By the
equation (\ref{en16000}), we know that for a perturbation of fixed
frequency , the bulk viscosity $\zeta$ is in proportion to the
reciprocal of the relaxation time $\tau$ under the low temperature
limit $1\ll \hat{\omega}\tau$ , that is $\zeta\propto\tau^{-1}$ ;
on the other hand $\zeta$ is in proportion to  $\tau$ under
condition of the high temperature limit $1\gg \hat{\omega}\tau$,
that is $\zeta\propto\tau$. Thereby, $\zeta(n_{B})$ could be
divided into two different behaviors, in another word,
$\zeta(n_{B})$ will increase with the temperature for low
temperature regime; and the circs is just opposite for the high
temperature regime. The transition temperature is the function of
baryon density and perturbation frequency (Figure
\ref{trantemperature} for $GPS2, g=2.0$ EOS).

Figure \ref{GPS fixed bulk viscosity} and Figure \ref{g fixed bulk
 viscosity} display the bulk viscosity of mixed phase as a function of baryon
 density for various temperatures and given EOSs. The numerical results show that i) the bulk viscosity of mixed
phase, either nucleon-hyperon phase or nucleon-quark phase or
nucleon-hyperon-quark phase, is of almost the same order of the
bulk viscosity of pure strange quark matter, ii) the differences
between EOSs only change the compositions and  sizes of the mixed
phase in the interior of the stars,  iii) the transition
temperature between low  and high temperature regime is about
$10^{8}\sim 10^{9} K$ in our models.

\section{R-mode instability windows of hybrid stars}
\begin{figure}
\centering
\includegraphics[width=0.6\textwidth]{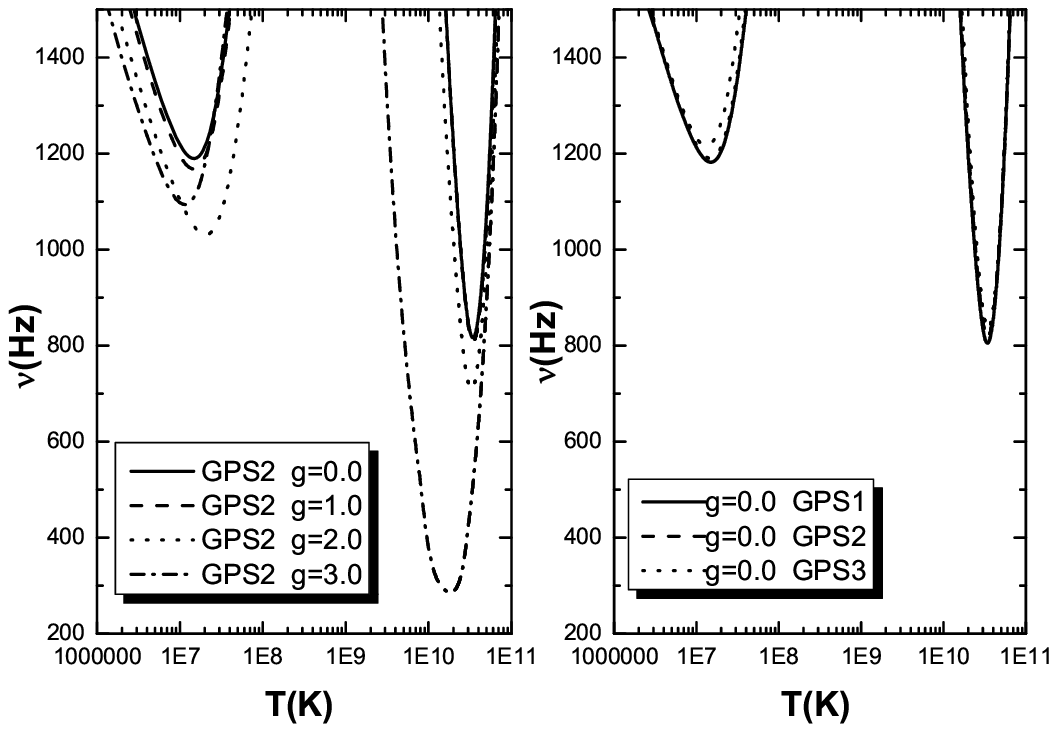}
\caption{The influences of medium effect and hadron EOSs on r-mode
instability windows. Each pair of curves represent the hybrid
stars of fixed maximum mass  discussed in the Figure
\ref{structure}(a) (b) respectively. Here $\nu$ is the critical
rotation frequency of the stars.}\label{rmode}
\end{figure}

The r-modes in a perfect fluid star succumbing  to gravitational
radiation (GR) drive the Chandrasekhar-Friedman-Schutz instability
for all rates of stellar rotation, and arise due to the action of
Coriolis force with positive feedback of the increasing of
GR(Andersson \& Kokkotas \cite{28}; Stergioulas \cite{29}).
Actually, the viscosity of stellar matter can hold back the growth
of the modes effectively. Based on the competition between the
destabilizing effect of GR and the damping effect of viscosity, a
r-mode instability window can be defined. Now, we could apply our
results of the bulk viscosity in the previous section together
with those of the HP and QP to considering the r-mode instability
windows of rotating compact stars using the standard formulae
given in literatures(Lindblom et al.\cite{30}; Lindblom et
al.\cite{31}). We first integrate the viscosity on the structure
of stars, which has been showed in Figure \ref{structure} for a
wide range of EOSs, to obtain the bulk viscosity timescale
$\tau_{B}$
\begin{equation}
\frac{1}{\tau_{B}}=-\frac{1}{2\widetilde{E}}(\frac{d\widetilde{E}}{dt})_{B},
\end{equation}
where
\begin{equation}
\widetilde{E}=\frac{1}{2}\alpha^{2}\Omega^{2}R^{-2}\int_{0}^{R}\epsilon(r)r^{6}dr,
\end{equation}
\begin{equation}
(\frac{d\widetilde{E}}{dt})_{B}=-4\pi\int_{0}^{R}\zeta(\epsilon(r))<|\overrightarrow{\nabla}\cdot\delta\overrightarrow{\upsilon}|^{2}>r^{2}dr.
\end{equation}
Here $\widetilde{E}$ is the mode energy in a frame rotating with
the star, $(\frac{d\widetilde{E}}{dt})_{B}$ is the rate at which
energy is  drained from the mode by viscosity, $\alpha$ is a
dimensionless amplitude coefficient, $\Omega$ is the angular
frequency of a spinning star, $\epsilon(r)$ is the energy density
of star at radius r, and $\delta\overrightarrow{\upsilon}$ is the
Eulerian velocity perturbation. In general the expansion of the
mode $\overrightarrow{\nabla}\cdot\delta\overrightarrow{\upsilon}$
is a complicated function of radius and angle. Although the
function can only been determined numerically, we adopt an
excellent analytical fit to those numerical solutions
\begin{equation}
<|\overrightarrow{\nabla}\cdot\delta\overrightarrow{\upsilon}|^{2}>={\frac{\alpha^{2}\Omega^{2}}{690}(\frac{r}{R})^{6}[1+0.86(\frac{r}{R})^{2}](\frac{\Omega^{2}}{\pi
G\bar{\epsilon}})^{2}},
\end{equation}
which is given by Lindblom (Lindblom \& Owen \cite{13}; Nayyar \&
Owen \cite{22}).
 Finally, we could
obtain the critical rotation frequency for the star with a solid
crust as a function of temperature following from
\begin{equation}
-\frac{1}{\tau_{GR}}+\frac{1}{\tau_{B}}+\frac{1}{\tau_{SR}}=0
\end{equation}
for low temperature limit($T< 10^{9}K$), and
\begin{equation}
-\frac{1}{\tau_{GR}}+\frac{1}{\tau_{B}}+\frac{1}{\tau_{mU}}=0
\end{equation}
for high temperature limit($T> 10^{9}K$). Here $\tau_{GR}$ is the
timescale for gravitational radiation to effect the
r-mode(Andersson \& Kokkotas \cite{28}), while $\tau_{SR}$ is the
timescale for surface rubbing due to the presence of the viscous
boundary layer if a solid crust exists(Bildsten \& Ushomirsky
\cite{32}; Andersson et al.\cite{33}), and $\tau_{mU}$ is the
viscosity timescale of modified URCA reactions for low density
hadronic matter(Lindblom et al.\cite{31}). Our results for the
critical rotation frequency  are shown in the Figure \ref{rmode}.

We may know from it that there are two instability windows for
such hybrid stars(Andersson \& Kokkotas \cite{28}). We find that
the r-mode instability is completely suppressed for the hybrid
stars from the low temperature windows. We realize that hadronic
EOSs, either stiffer of softer, hardly influence the instability
windows. It is because that the structure of the star including
quark phase hardly changes for different $GPS$ EOSs. The
increasing medium effect of quarks enlarges the window at low $g$
values but reduces the window at high $g$ values. The properties
may imply that extremely rapid rotating or submillisecond pulsars
would exist, and just be the hybrid stars.

\section{Conclusion}
Hyperons and quarks exist  in cores of  neutron stars at high
density. To evaluate accurately the bulk viscosity (due to
strangeness-changing four-baryon weak interactions involving both
hyperons and quarks non-leptonic processes), it is necessary to
compute detailed and accurate models for the compositions and
structures of the neutron-star cores. We use $GPS$ EOSs for
hadrons and $SGT$ model with medium effect for quarks to evaluate
composition of the nuclear matter inside the stellar core and
solve  the TOV equation to obtain the stellar structure, and
particularly study the effect of differences between EOSs on the
structures of stars.

We numerically calculate the bulk viscosity of mixed phase matter
of various EOSs, and especially give the treatment of the case
that  involves both hyperons and quarks reactions. It is a
generalization of single reaction case in previous work (Drago et
al.\cite{21}). Our results show that there is high coefficient of
bulk viscosity as long as exotic particle appears in high density
nuclear matter, either strange quark or hyperon.

We have considered medium effect of quark matter. The bulk
viscosity of hybrid star increases as the coupling among quarks
becomes stronger. The coefficient of bulk viscosity at large
coupling constant is higher than assumed in reference (Drago et
al.\cite{21}), and the mixed phase regime is expanded. Just so,
the hybrid stars have extremely high limiting rotation frequencies
from the low temperature windows even if the stellar masses are
large. The submillisecond  pulsars may exist.

We must emphasize that we are concerned about the bulk viscosity
of mixed normal nucleon-hyperon-quark phase. Nucleon, hyperon
superfluidity and quark superconductivity may occur at $T < 10^{8}
K$ and given critical density. These effects would significantly
influence the dissipation of hybrid star matter. These
considerations, which are made in hyperon star model (Lindblom \&
Owen \cite{13}) and hybrid star model (Drago et al.\cite{21}), are
our future work.

\section*{Acknowledgments}
 This work is supported by NFSC under Grant Nos.90303007 and
10373007, and the Ministry of Education of China with project
No.704035. We are especially grateful to Professor D.F. Hou for
reading and checking the text.

\clearpage

\begin{table*}
\begin{center}
\begin{tabular}{|c|c|c|c|c|c|c|}\hline
$Name$ & $\rho_{0}(fm^{-3})$ & $B/A(Mev)$ & $a_{sym}(Mev)$&
$K(Mev)$ & $m^{*}/m_{N}$ & $x_{\sigma}=x_{\rho}=x_{\omega}$
\\ \hline
$GPS1$ & 0.150 & -16.0 & 32.5 & 250 & 0.83 &
 $\sqrt{\frac{2}{3}}$ \\ \hline
$GPS2$ & 0.150 & -16.0 & 32.5 & 300 & 0.83 &
 $\sqrt{\frac{2}{3}}$ \\ \hline
$GPS3$ & 0.150 & -16.0 & 32.5 & 350 & 0.83 &
 $\sqrt{\frac{2}{3}}$ \\\hline
\end{tabular}
\end{center}
\vspace{5mm} \caption{The nuclear matter properties of the $GPS$
equations of state. $\rho_{0}$ is the saturation density, $B/A$ is
the binding energy, and the incompressibility,  effective mass and
the symmetry energy are denoted by $K$, $m^{*}/m_{N}$ and
$a_{sym}$ respectively. Here we choose hyperon coupling parameters
$x_{\sigma}$, $x_{\rho}$, $x_{\omega}$ to be $\sqrt{\frac{2}{3}}$
times the nucleon-meson coupling parameters. }. \label{parameter}
\end{table*}

\begin{figure}
\includegraphics[width=0.9\textwidth]{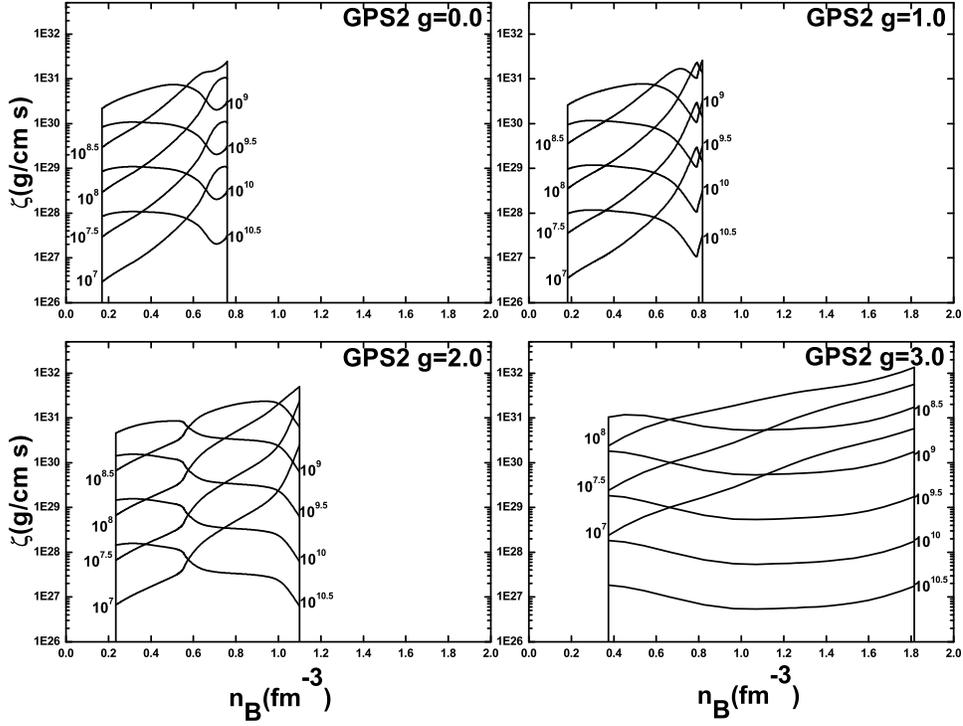}
\caption{Influences of medium effect on bulk viscosities  as
functions of  baryon number densities for a range of temperatures.
Parameters are the same with Figure 1.} \label{GPS fixed bulk
viscosity}
\end{figure}

\begin{figure}
\centering
\includegraphics[width=0.6\textwidth]{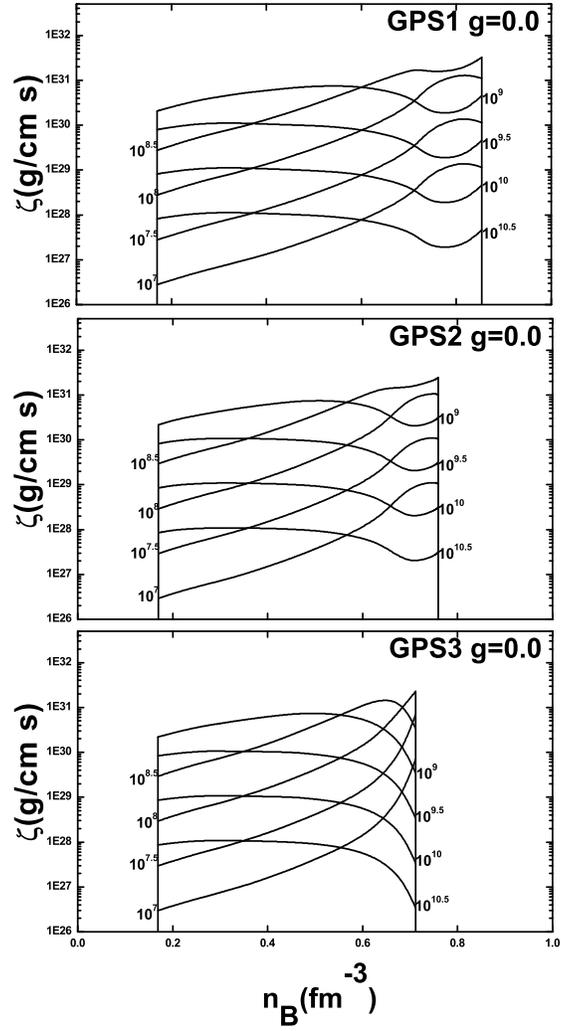}
\caption{Influences of hadron EOSs on bulk viscosities as
functions of  baryon number densities for a range of temperatures.
Parameters are the same with Figure 2.} \label{g fixed bulk
viscosity}
\end{figure}

\begin{thebibliography}{}
\bibitem[1998]{17}Andersson,N. Astrophys.J, \textbf{502} 708 (1998).
\bibitem[2000]{33}Andersson,N., Jones,D.J. and Kokkotas,K.D. et al., Astrophys.J, \textbf{534} L75-78 (2000).
\bibitem[2001]{28}Andersson,N., Kokkotas,K.D. Int.J.Mod.Phys, \textbf{10} 381 (2001).
\bibitem[1971]{25}Baym,G., Pethick,C., Sutherland,P. Astrophys.J, \textbf{170} 299 (1971).
\bibitem[2000]{32}Bildsten,L., Ushomirsky,G. Astrophys.J, \textbf{529} L33 (2000).
\bibitem[1996]{27}Dai,Z.G., Lu,T. Z.Phys.A, \textbf{355} 415-420 (1996).
\bibitem[2005]{21}Drago,A., Lavagno,A., Pagliara,G. Phys.Rev.D, \textbf{71} 103004 (2005).
\bibitem[1998]{18}Friedman,J.L., Morsink,S.M. Astrophys.J, \textbf{502} 714 (1998).
\bibitem[1995]{23}Ghosh,S.K., Phatak,S.C., Sahy,P.K. Z.Phys.A, \textbf{352} 457 (1995).
\bibitem[1997]{19}Glendenning,N.K. \textit{Compact stars} (Springer-Verlag, 1997).
\bibitem[1992]{20}Glendenning,N.K. Phys.Rev.D, \textbf{46} 1247 (1992).
\bibitem[1992]{5}Glendenning,N.K., Weber,F., Moszkowski,S.A. Phys.Rev.C, \textbf{45},844 (1992).
\bibitem[1994]{11}Goyal,A., Gupta,V.K., Pragya and Anand,J.D. Z.Phys.A, \textbf{349}
93 (1994).
\bibitem[2000]{12}Haensel,P., Levenfish,K.P. Astron.Astrophys, \textbf{357} 1157 (2000).
\bibitem[2002]{34}Haensel,P., Levenfish,K.P.,
Yakovlev,D.G. Astron.Astrophys, \textbf{381} 1080 (2002).
\bibitem[1992]{3}Haensel,P., Schaeffer,R. Phys.Rev.D, \textbf{45},4708 (1992).
\bibitem[2001a,b]{6}Jones,P.B. Phys.Rev.Lett, \textbf{86},1384 (2001a).
\bibitem[2001]{7}Jones,P.B. Phys.Rev.D, \textbf{64}, 084003 (2001b).
\bibitem[1999]{26}Landau,L.D. and Lifshitz,E.M. \textit{Fluid Mechanics} 2nd edition (Butterworth-Heinemann,Oxford, 1999).
\bibitem[1999]{31}Lindblom,L., Mendell,G., Owen,B.J. Phys.Rev.D, \textbf{60} 064006 (1999).
\bibitem[2002]{13}Lindblom,L., Owen,B.J. Phys.Rev.D, \textbf{65} 063006 (2002).
\bibitem[1998]{30}Lindblom,L., Owen,B.J., Morsink,S.M. Phys.Rev.Lett, \textbf{80} 4843 (1998).
\bibitem[1992]{10}Madsen,J. Phys.Rev.D, \textbf{46} 3290 (1992).
\bibitem[2005]{22}Nayyar,M., Owen,B.J. astro-ph/0512041.
\bibitem[1989a]{2}Sawyer,R.F. Phys.Rev.D, \textbf{39}, 3804 (1989a).
\bibitem[1989b]{9}Sawyer,R.F. Phys.Lett, \textbf{B233}, 412 (1989b).
\bibitem[1979]{1}Sawyer,R.F., Soui,A. Astrophys.J, \textbf{230}, 859
(1979).
\bibitem[1997]{24}Schertler,K., Greiner,C., Thoma,M.H. Nucl.Phys.A, \textbf{616} 659 (1997).
\bibitem[2003]{29}Stergioulas,N. Living.Rev.Rel, \textbf{6} 3 (2003), and references therein, gr-qc/0302034.
\bibitem[1989]{35}Taylor,H., Weisberg,J.M. Astrophys.J, \textbf{345}, 434 (1989).
\bibitem[1984]{8}Wang,Q.D., Lu,T. Phys.Lett,
\textbf{B148},211 (1984).
\bibitem[2005]{4}Weber,F. \textit{Progress in Particle and Nuclear Physics},
54(1),193-288 {2005}.
\bibitem[2005]{16}Zheng,X.P., Kang,M., Liu,X.W., Yang,S.H. Phys.Rev.C, \textbf{72} 025809 (2005).
\bibitem[2004]{15}Zheng,X.P., Liu,X.W., Kang,M., Yang,S.H. Phys.Rev.C, \textbf{70} 015803 (2004).
\bibitem[2003]{14}Zheng,X.P., Yang,S.H., Li,J.R. Astrophys.J, \textbf{585} L135 (2003).


\end{thebibliography}
\end{document}